\documentclass{elsart}

\usepackage{epsfig}

\usepackage{amssymb}

\begin{document}

\begin{frontmatter}



\title{High Energy Beam Test of the PHENIX Lead--Scintillator EM Calorimeter}


\author[ORNL]{T.C. Awes}
\author[RBRC]{A. Bazilevsky}
\author[Muenster]{S. Bathe}
\author[Muenster]{D. Bucher}
\author[Muenster]{H. Buesching}
\author[RBRC]{Y. Goto}
\author[Kyoto]{K. Imai}
\author[Kurchatov]{M.S. Ippolitov}
\author[BNL]{E. Kistenev}
\author[Muenster]{V. Mexner}
\author[Muenster]{T. Peitzmann}
\author[RIKEN,RBRC]{N. Saito}
\author[Kyoto,RIKEN]{H. Torii}
\author[BNL]{S.N. White}

\address[BNL]{\sl Brookhaven National Laboratory, Upton, NY 11973-5000, USA}
\address[Kurchatov]{\sl Russian Research Center ``Kurchatov Institute'', 
Moscow, Russia}
\address[Kyoto]{\sl Kyoto University, Kyoto 606-8502, Japan}
\address[Muenster]{\sl Institut fuer Kernphysik, University of Muenster, 
D-48149 Muenster, Germany}
\address[ORNL]{\sl Oak Ridge National Laboratory, Oak Ridge, TN 37831, USA}
\address[RIKEN]{\sl RIKEN, Wako, Saitama 351-0198, Japan}
\address[RBRC]{\sl RIKEN BNL Research Center, Brookhaven National Laboratory, 
Upton, NY 11973-5000, USA}

\begin{abstract}
In the PHENIX experiment at RHIC, the electro-magnetic calorimeter plays an important role in both the
heavy-ion and spin physics programs for which it was designed.
In order to measure its performance in the energy range up to 80GeV, a beam
test was performed at the CERN-SPS H6 beam line.
We describe the beam test and present results on calorimeter performance with
pion and electron beams.
\end{abstract}

\end{frontmatter}


\section{Introduction}\label{chap:intro}
The PHENIX experiment at RHIC started data taking in 2000, with relativistic
heavy-ion collisions. Subsequent runs will also use polarized proton beams to
carry out a program of spin physics.
In PHENIX, the electro-magnetic (EM) calorimeter is the primary tool for measuring 
photons and electrons/positrons.
In order to cover topics in both physics programs, e.g. a thermal photon
measurement in relativistic heavy-ion collisions, and prompt photon, $\pi^0$ and
weak boson measurements in spin physics, the EM calorimeter needs to
cover a wide energy range extending from a few hundred MeV to 80GeV.
A goal of the spin physics program is to measure differential cross sections
of prompt photon and $\pi^0$'s to an accuracy of 10\%.
A 2\% accuracy in the calorimeter energy scale is required to achieve this for
$p_T >$ 10GeV/$c$ of interest, because the cross sections fall steeply as
their energy increases.
The EM calorimeter was originally designed for relativistic heavy-ion physics.
There are two kinds of calorimeter in the PHENIX detector.
One is a Shashlik \cite{SNW94,JB_NIM94,ET_NIM96} type lead--scintillator sampling
calorimeter (PbSc) and another is a lead glass calorimeter (PbGl).
Table \ref{tbl:emcal} shows their basic parameters.
A ``super-module'' is composed of 12$\times$12 channels for the PbSc
calorimeter and 4$\times$6 channels for the PbGl calorimeter.
The total EM calorimeter system in the PHENIX detector consists of
the PbSc super-modules and the PbGl super-modules.

\begin{table}[hbtp]
\begin{center}
\begin{tabular}{lll|c|c}
\hline
\hline
 &	&               & \makebox[1.5cm]{PbSc} &\makebox[1.5cm]{PbGl} \\
\hline
 & radiation length ($X_0$)
        & [mm]          & 21                    & 29                   \\
 & Moliere radius
	& [mm]          & $\sim$30              & 37                   \\
\hline
\multicolumn{3}{l|}{channel}
	                &                       &                      \\
 & cross section
	& [mm$^2$]      & 52.5$\times$52.5      & 40$\times$40         \\
 & depth
	& [mm]          & 375                   & 400                  \\
 &	& [$X_0$]       & 18                    & 14                   \\
 & \multicolumn{2}{l|}{$\eta$ coverage}
                        & 0.011                 & 0.008                \\
 & \multicolumn{2}{l|}{$\phi$ coverage}
                        & 0.011                 & 0.008                \\
\hline
\multicolumn{3}{l|}{super-module}
	                &                       &                      \\
 & \multicolumn{2}{l|}{number of channels}
	                & 144 (12$\times$12)    & 24 (4$\times$6)      \\
\hline
\multicolumn{3}{l|}{sector}
	                &                       &                      \\
 & \multicolumn{2}{l|}{number of super-modules}
	                & 18 (3$\times$6)       & 192 (12$\times$16)   \\
\hline
\multicolumn{3}{l|}{total system}
	                &                       &                      \\
 & \multicolumn{2}{l|}{number of sectors}
	                & 6                     & 2                    \\
 & \multicolumn{2}{l|}{number of channels}
	                & 15552                 & 9216                 \\
 & \multicolumn{2}{l|}{$\eta$ coverage}
                        & 0.7                   & 0.7                  \\
 & \multicolumn{2}{l|}{$\phi$ coverage}
                        & 90$^\circ$+45$^\circ$ & 45$^\circ$           \\
\hline
\hline
\end{tabular}
\caption{
Basic parameters of two kinds of PHENIX EM calorimeter.
}
\label{tbl:emcal}
\end{center}
\end{table}

The calorimeter's energy resolution, linearity  and
hadron rejection had already been measured at BNL-AGS in the energy range up to
7GeV \cite{GD_IT98a}.
In order to extend these measurements to the energy range up to 80GeV, a beam
test was performed at the CERN-SPS H6 beam line in 1998.

In this article, we describe the beam test and present results of the PbSc data
analysis.

\section{Setup}\label{chap:setup}

\begin{figure}[hbtp]
\begin{center}
\psfig{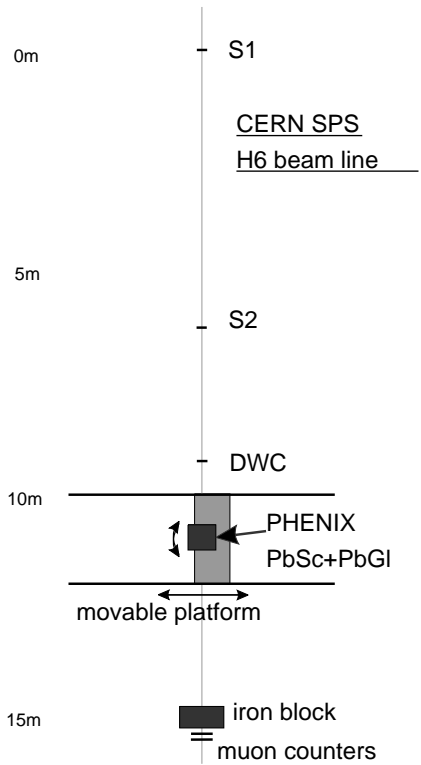}
\caption{Setup of the high energy beam test at CERN.}
\label{fig:setup}
\end{center}
\end{figure}

Figure \ref{fig:setup} shows a setup of the beam test.
One PbSc super-module and four PbGl super-modules were located at the
H6 beam line, and both were tested with electron beams in the momentum range of
10GeV/$c$ to 80GeV/$c$ and $\pi^+$ beams of 40GeV/$c$.
Both kinds of calorimeter were placed on a movable platform to change
positions and angles of the incident beam on the calorimeter.
Delay-line Wire Chambers (DWC) \cite{DWC} were located just in front of the
calorimeter for measurements of the vertical and horizontal beam incident
position.
Incident-position dependence of the energy deposit were measured and
corrected using the DWC.
Two scintillators (S1 and S2) were used as trigger counters, and 
two other scintillators (muon counters) were set behind iron blocks to identify muons
in the beam.
There was a \v Cerenkov Differential counter with Achromatic Ring focus (CEDAR)
further upstream of the S1 for electron identification.

We used the 10GeV/$c$ muon beams for channel-by-channel gain adjustment of
the PbSc super-module channels in addition to the electron beams.
For time-dependent gain drift correction of the PbSc calorimeter, we used a
laser monitoring system \cite{GD_IT98b}.

The DWC has good position resolution (0.2mm) and high single-particle detection
capability (2$\times$10$^5$ particles/sec).
It consists of one anode-wire plane and two cathode wire planes which
surround the anode plane.
The cathode planes have 2mm wire spacing.
Their wires are connected with a delay-line through which signals are read out
by a TDC module.
Timing information corresponds linearly to position information.
The active area is 100$\times$100 mm$^2$, and the area with
linear response is 80$\times$80 mm$^2$

A beam trigger was composed of a S1 signal (``$S1$''), a S2 signal (``$S2$'')
and a coincident signal of two muon counters (``$\mu$'').
An electron trigger was made by $S1 \otimes S2 \otimes \bar{\mu}$
and a muon trigger was made by $S1 \otimes S2 \otimes \mu$.
Triggers for pedestal measurement and for the laser monitoring system
were used to take those data between beam spills.

We used two different HV settings;
\begin{itemize}
\item
a "normal" HV setting (1.23--1.29kV) for energy measurements up to 80GeV
\item
a "low" HV setting (1.13--1.19kV) for energy measurements up to 160GeV.
\end{itemize}

To readout PMT signals from the calorimeter, we used front-end electronics (FEE)
from the CERN experiment WA98\cite{ALW94}.

\section{Analysis}\label{chap:analysis}

The deposited energy in each channel was calculated by multiplying the ADC count
by a calibration factor, $C(t)$ (GeV/count).
The calibration factor has time dependence.
We parameterize the time dependence by an initial gain factor, $G$ (GeV/count),
and a gain drift, $D(t)$.
\begin{displaymath}
C(t) = G \times D(t).
\end{displaymath}
The time dependent factor, $D(t)$, is defined to be 1 at the time of the muon calibration
run.

The ADC count is derived from low and high gain ADCs which
are both implemented to cover  a wider dynamic range.
We found a 1\% difference between the ratio of high-to-low gain
in electron trigger events and in laser trigger events.
To determine the channel dependent high-to-low gain ratio, we use the measured
ratio in electron beam trigger events for the 40 channels which have hits in the electron
beam data.
For the other channels, we measure the ratio with the laser trigger events.
These channels have only a small contribution to the electron beam energy measurement
hence a small effect on the systematic error.
The average value of the high-to-low ratio is 7.8.
The ADC value is
derived from the low gain ADC when the low gain ADC count is larger than
90 counts.
We also found there is no time dependence of the high-to-low ratio during whole
run.
The systematic error caused by the high-to-low ratio is less than
1\% for those channels which have the electron beam data.
There is a negligible contribution of the other channels to the systematic
error.

The gain factor is adjusted channel-by-channel by using muon trigger
events.
In order to identify muons, the following selections are applied.

\begin{enumerate}
\item
The channel which has the largest energy deposit
in all channels must have more than 80\% of the energy sum of all channels.
\item The number of channels which have energy deposit
more than 130MeV must be zero or one.
\end{enumerate}

When a muon beam penetrates one channel longitudinally, the most probable energy
deposit is about 300MeV.
Selection 1 requires that there are some hits which make a peak on that
channel.
Selection 2 requires it is a minimum-ionizing single particle and rejects
background from other kinds of particles, and multi-hit of particles.
By requiring both selections, muons which penetrate one channel are
selected.

After the muon selection,
we have more than 100 muons in each of the 40 channels.
We adjust gains of these 40 channels so that the  MIP peak position is at the same
energy.
The peak position is determined to a precision of 2--3\%.
In order to improve precision of the gain adjustment and to obtain the gain
factor for the other channels, we use electron beam trigger events.
The remaining errors of channel-by-channel adjustment of the gain factor
is 3\% in total.
These errors
are statistical ones. The systematic errors are smaller than these.

An absolute value of the gain factor to provide correct energy scale
is fixed at the electron beam energy of 20GeV.
The result shows the average value of the gain factor is 110 (count/GeV).

The time dependence of the gain was obtained
using laser trigger events.
The time variation of the laser amplitude was less than 3\%
reflecting the stability of the laser output.
To monitor the fluctuation of the laser output, we use a truncated mean of
the laser amplitude.

The gain drift factor obtained with this method works reliably over
periods of order a few hours.
Between some sets of runs the gain drift is normalized by using the beam
energy of 20GeV in the period.
The accuracy of the beam energy is 1\% at 15GeV\cite{GA_EPJ99}.

In this analysis, the total deposited energy is defined as a sum of energies
in the $5 \times 5$ channels centered on the channel with the maximum energy deposit.
The total energy is corrected by a position dependent factor.

\begin{figure}[htbp]
\begin{center}
\psfig{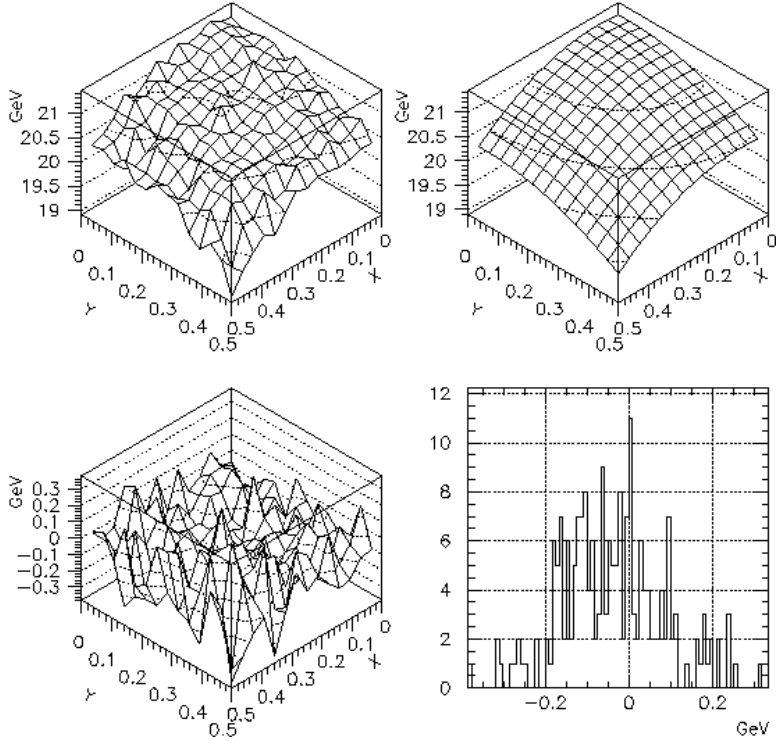}
\caption{Upper-left: Hit position dependence of the $5\times5$ energy sum for
the 20GeV electron beam, where $(X,Y)$ shows a hit position obtained by the
DWC.
Upper-right: Fitted hit position dependence.
Lower-left: Deviation of the energy sum from the fitted position dependence
Lower-right: Projection of the deviation.}
\label{fig:posdep_syserr}
\end{center}
\end{figure}

The upper-left figure of Fig.\ref{fig:posdep_syserr} shows the hit position
dependence of the energy sum in $5 \times 5$ channels for the 20GeV
electron beam.
In this figure, the coordinate $(X,Y)$ shows a hit position in one channel
obtained by the DWC.
The hit position $(0,0)$ presents the center of the channel and
$(1,1)$ presents the edge of the channel.
The position dependence is fitted as shown in the upper-right figure of
Fig.\ref{fig:posdep_syserr} by the following formula.
\begin{equation}
1 + a \times (X^2+Y^2) + b \times (X^4+Y^4) + c \times X^2 \cdot Y^2
\label{eqn:posdep}
\end{equation}
We obtained a best fit with the following parameters;
\begin{eqnarray}
a &=& -0.3079	\nonumber \\
b &=& 0.3643	\nonumber \\
c &=& -0.02894	\nonumber
\end{eqnarray}

We use these parameters to correct for the position dependence of the energy
measurement.
The lower-left figure shows the deviation of the energy sum from the fitted
hit position dependence and the lower-right figure shows a projection of the
deviation.
The deviation is 0.5\% of the energy sum.
The systematic error remaining after the position dependence correction is
evaluated to be 0.5\%.

\section{Results}\label{chap:results}

\begin{figure}[htbp]
\begin{center}
\psfig{file=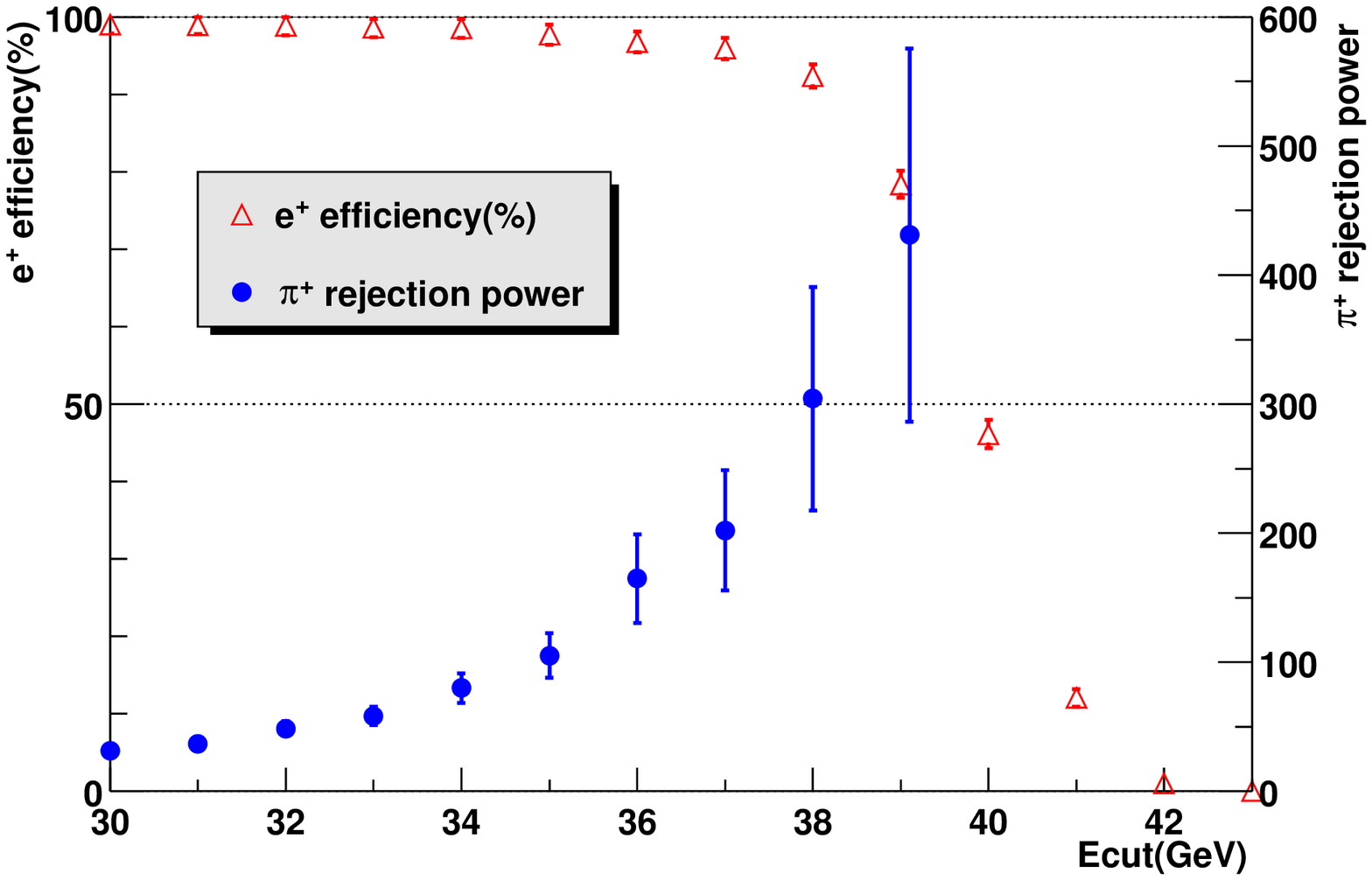,height=10.0cm}
\caption{The efficiency for the 40GeV positron beam when we require a measured energy deposit greater than $E_{cut}$,
and the pion rejection power for the 40GeV $\pi^{+}$ beam obtained
with the same cut.}
\label{fig:pirej}
\end{center}
\end{figure}

Figure \ref{fig:pirej} shows the efficiency for the 40GeV positron beam when we 
require a measured energy deposit greater than  $E_{cut}$, and the pion rejection power for the 
40GeV $\pi^+$ beam obtained with  the same cut.

The energy resolution at each energy point was obtained by fitting a Gaussian
distribution within $\pm 2 \sigma$ around the electron peak.
The electron trigger events contained a 10\% pion contamination.
The contamination in the electron peak region
is less than 0.1\% because only 1\% of the pions deposit more than
90\% energy in the calorimeter.
The $\chi^2$ of the fitting is reasonably small.

\begin{figure}[htbp]
\begin{center}
\psfig{file=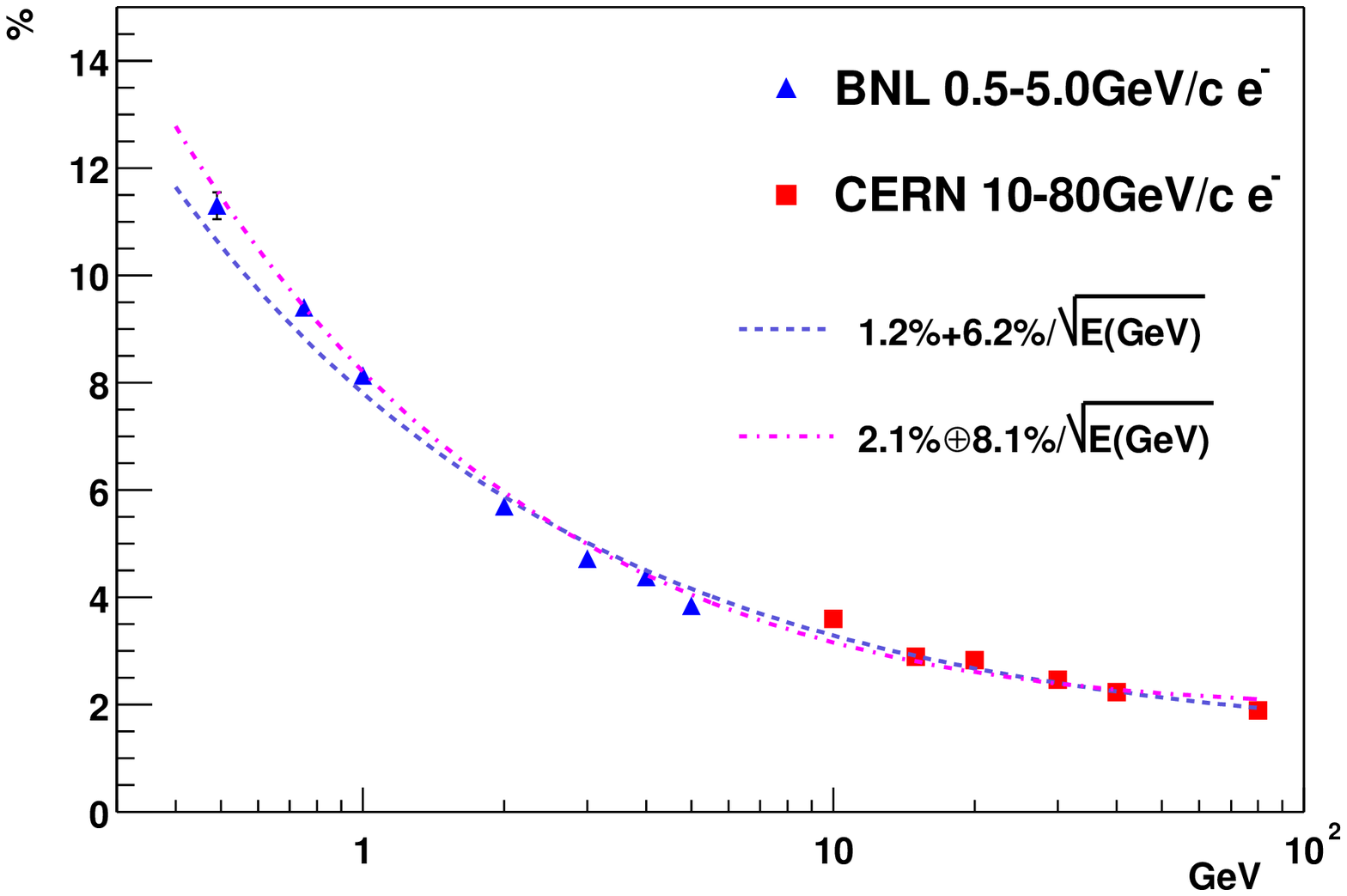,height=10.0cm,width=12.0cm}
\caption{Energy resolution obtained by both beam tests at
BNL and CERN.
A dashed line shows the result of fitting by a linear formula, $\sigma_{E}/E =
1.2\% + 6.2\% / \sqrt{E(GeV)}$.
A dashed dotted line shows the result of fitting by a quadratic formula,
$\sigma_{E}/E = 2.1\% \oplus 8.1\% / \sqrt{E(GeV)}$.}
\label{fig:ereso}
\end{center}
\end{figure}

Figure \ref{fig:ereso} shows energy resolution obtained by both
beam tests at CERN and BNL.
They can be fit with linear or quadratic expressions. Only statistical errors are taken into
 account in the fits.
We estimate an additional 1\% systematic error based on the reproducibility of the measurements
at each energy point.
The results of the fits are
\begin{displaymath}
\sigma_{E}/E = 1.2\% + \frac{6.2\%}{\sqrt{E(GeV)}}
\end{displaymath}
\begin{displaymath}
\sigma_{E}/E = 2.1\% \oplus \frac{8.1\%}{\sqrt{E(GeV)}}
\end{displaymath}
where $\oplus$ denotes a square of the quadratic sum, $\alpha\oplus\beta
= \sqrt{\alpha^2+\beta^2}$.
They are valid in the energy region of 0.5GeV to 80GeV with 1\% systematic
uncertainty.

\begin{figure}[htbp]
\begin{center}
\psfig{file=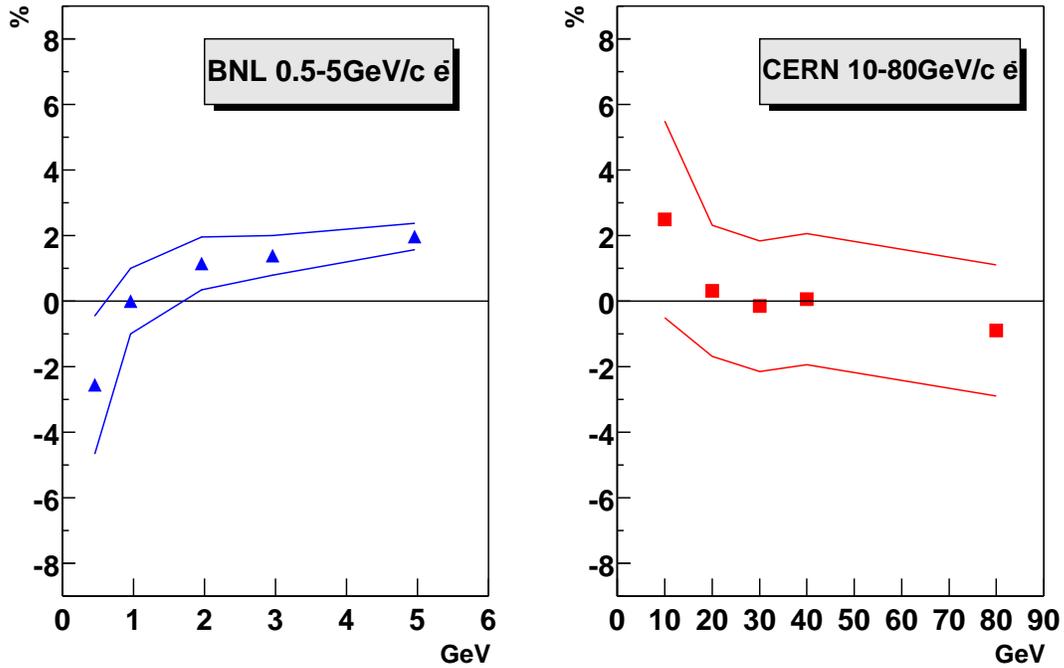,height=10.0cm}
\caption{Linearity of the $5\times5$ energy sum for both beam tests at BNL
(left) and CERN (right).
Solid lines show total systematic uncertainties in the analysis.
Absolute level of two beam test is not normalized.}
\label{fig:Dst_lin}
\end{center}
\end{figure}

Figure \ref{fig:Dst_lin} shows the residual (measured energy with the calorimeter less
the beam energy, divided by the beam energy) of the energy sum in $5 \times 5$
channels versus the beam energy.
We see that the calorimeter response is linear within 2\% systematic uncertainties in the
energy region from 20GeV to 80GeV.
There is some indication of a 2\% deviation from linearity at 10GeV, however this is within our systematic
errors.
Such a deviation cannot be due to one of the following corrections;
the gain drift correction, pedestal subtraction, high-to-low ratio
correction, or the energy sum of $5\times5$ channels.
We considered the following possible sources;
\begin{itemize}
\item
run-time problems in the monitoring system
\item
linearity of the WA98 FEE
\item
linearity of the PMT
\item
an inherent non-linearity in the calorimeter due for example to the interplay between light 
attenuation in the wave-length-shifter (WLS) readout fibers and
longitudinal shower leakage beyond the calorimeter. 
\end{itemize}
Linearity of the WA98 electronics which was used to digitize signals from the
calorimeter was investigated. We found that it has linear response within 1\%.
Linearity of the PMT had also been investigated and confirmed to have linear
response within 2\% \cite{GD_IT96}.
The light leakage in the WLS fiber is evaluated by the simulation program.
The energy leakage is evaluated to be 1\% at 10GeV and 4\% at 80GeV
\cite{DM_PR72,GB_NP70}.
These effects tend to cancel one another, so we do not expect the intrinsic 
non-linearity to be as large as 2\%. In summary, the only place where
the measured residual approaches the
limit of 2\% is at 10 GeV. Nevertheless, the residual is consistent within our systematic error
of being linear at this energy. We've examined a number of possible causes for a deviation from
linearity but none has been found.  

Position resolution of the beam hit position is evaluated with the logarithmic
method \cite{TCA_NIM92}.
In the logarithmic method, the position is determined by the following formula.
\begin{displaymath}
X = \frac{\displaystyle{\sum_{i=1 \ldots N} C_{i} \times x_{i}}}
{\displaystyle{\sum_{i=1 \ldots N}C_{i}}}
\end{displaymath}
where $x_{i}$ denotes a center of each channel in the horizontal direction.
Similarly, $Y$ is defined in the vertical direction.
The weights $C_{i}$ are;
\begin{eqnarray}
R_{i} &=& Max[ 0, E_{i}/E_{total} ]  \nonumber \\
C_{i} &=& Max[ 0, log(R_{i}) + C_{0} ] \nonumber
\end{eqnarray}
where $E_{total}$ is the total energy, $E_{total} =
\displaystyle{\sum_{i=1 \ldots N} E_{i}}$ and $C_0$ is a constant.
A larger value of $\alpha$ and $\beta$ is expressed by $Max[\alpha,\beta]$.

The deviation in a short period shows that systematic uncertainty in the logarithmic
method is 2mm.

\begin{figure}[htbp]
\begin{center}
\psfig{file=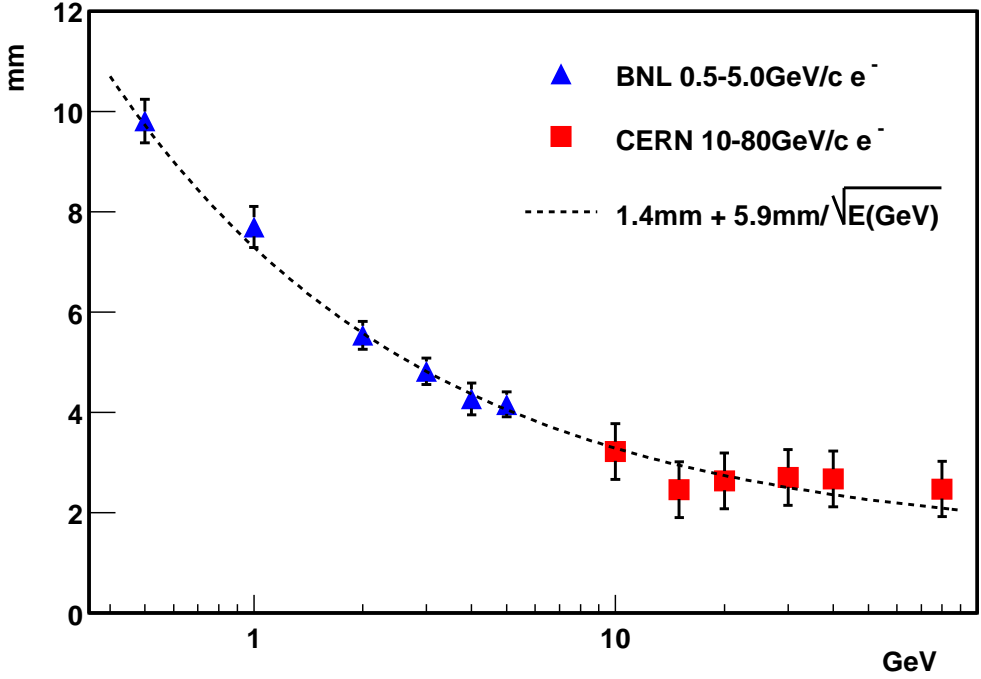,height=10.0cm}
\caption{Position resolution obtained by both beam tests at BNL and CERN.
A dashed line shows the result of fitting, 1.4 mm +
5.9 mm / $\sqrt{E(GeV)}$.}
\label{fig:posreso}
\end{center}
\end{figure}

Figure \ref{fig:posreso} shows the position resolution obtained by both beam
tests at CERN and BNL.
The points can be fitted by a formula;
\begin{displaymath}
\sigma_{x}(\mbox{mm}) 
= 1.4(\mbox{mm})+\frac{5.9(\mbox{mm})}{\sqrt{E(\mbox{GeV})}}.
\end{displaymath}

\section{Conclusion}\label{chap:summary}

We measured the energy resolution, linearity and position
resolution of the PHENIX EM calorimeter in a test beam at
CERN.
For the PbSc calorimeter, we obtained energy resolution of;
\begin{displaymath}
\sigma_{E}/E = 1.2\% + \frac{6.2\%}{\sqrt{E(GeV)}}
\end{displaymath}
\begin{displaymath}
\sigma_{E}/E = 2.1\% \oplus \frac{8.1\%}{\sqrt{E(GeV)}}
\end{displaymath}
and position resolution;
\begin{displaymath}
\sigma_{x}(\mbox{mm}) 
= 1.4(\mbox{mm})+\frac{5.9(\mbox{mm})}{\sqrt{E(\mbox{GeV})}}.
\end{displaymath}

A major purpose of the test was to investigate the performance of
the calorimeter in the energy range up to 80GeV and, in particular, the linearity
of response versus beam energy. Since our goal in PHENIX is to measure
prompt photon and $\pi^0$  production cross sections with the calorimeter within 10\% errors
it is important to understand the linearity of the calorimeter at the level of 2\%.
In the analysis of the PbSc calorimeter, the response is found to be linear to approximately 
this level.

\section*{Acknowledgment}
	We are grateful for test beam support we received while at CERN from 
Konrad Elsener and Alfredo Placci. We would also like to thank the ALICE-Phos group for 
assistance during the run and our colleagues in the PHENIX EM Calorimeter group for discussions.
We also acknowledge the support of Hideto En'yo and Bill Zajc.
This work has been
partially supported within the framework of the RIKEN--BNL Collaboration for RHIC Spin
Physics and under a grant from the US-Dept. of Energy.

\end{document}